\pdfoutput=1

\documentclass[preprint,showpacs,preprintnumbers,amsmath,amssymb,floatfix,endfloats*]{revtex4}

\usepackage{graphicx}
\usepackage{dcolumn}
\usepackage{bm}
\usepackage{amsfonts}

\DeclareMathAlphabet{\mathsfsl}{OT1}{cmr}{bx}{it}
\begin{document}
\title{Structural relaxation of porous glasses due to internal stresses and
deformation under tensile loading at constant pressure}
\author{Nikolai V. Priezjev$^{1,2}$ and Maxim A. Makeev$^{3}$}
\affiliation{$^{1}$Department of Mechanical and Materials
Engineering, Wright State University, Dayton, OH 45435}
\affiliation{$^{2}$National Research University Higher School of
Economics, Moscow 101000, Russia}
\affiliation{$^{3}$Department of Chemistry, University of
Missouri-Columbia, Columbia, MO 65211}
\date{\today}
\begin{abstract}

The time evolution of the pore size distributions and mechanical
properties of amorphous solids at constant pressure is studied using
molecular dynamics simulations. The porous glasses were initially
prepared at constant volume conditions via a rapid thermal quench
from the liquid state to the glassy region and allowing for
simultaneous phase separation and material solidification. We found
that at constant pressure and low temperature, the porous network
becomes more compact and the glassy systems relocate to
progressively lower levels of the potential energy. Although the
elastic modulus and the average glass density both increase with the
waiting time, their dependence is described by the power-law
function with the same exponent. Moreover, the results of numerical
simulations demonstrated that under tensile loading at constant
pressure, low-density porous samples become significantly deformed
and break up into separate domains at high strain, while dense
glasses form a nearly homogeneous solid material.

\end{abstract}

\pacs{34.20.Cf, 68.35.Ct, 81.05.Kf, 83.10.Rs}


\maketitle

\section{Introduction}

Understanding the influence of pore morphology and porosity on
mechanical and physical properties of metallic porous materials is
important for structural applications, including wear-resistant
tools and biomedical implants, as well as functional applications in
catalysis and heat conduction~\cite{Fu18,Xie16}. A number of
fabrication techniques including phase separation and additive
manufacturing open the possibility of obtaining porous structures
with various pore sizes, shapes and distributions~\cite{Fu18}.
Recent experimental and computational studies have demonstrated that
plastic deformation of bulk metallic glasses with a regular array of
pores is controlled by shear localization between neighboring
pores~\cite{Bargmann14,EckertMet15,EckertAct16}. Moreover, detailed
atomistic and continuum simulations have shown that void defects
result in shear band nucleation in metallic glasses at strain lower
than in uniform samples, and the critical strain depends on the
shape and size of the defects~\cite{Gouripriya17,Luo18}.  It was
also found that tensile plasticity of metallic glass matrix
composites can be improved by increasing particle volume fraction,
which provides the resistance to shear band
propagation~\cite{Jiang17}. Similarly, the plasticity of metallic
glasses under tensile loading can be enhanced by introducing an
array of crystalline particles or by increasing
temperature~\cite{Song17}. However, the exact relationships between
elastic modulus, tensile strength and pore size, shape and
distribution have yet to be established.

\vskip 0.05in

Using molecular dynamics simulations, it was recently shown that
rapid isochoric quenching of a glass-forming mixture from a high
temperature liquid state across the glass transition leads to the
formation of porous glassy media~\cite{Kob11,Kob14,Makeev18}. The
kinetics of phase separation and solidification of the glass phase
generally depend on the average glass density as well as the rate
and depth of thermal quench~\cite{Kob11}.   In particular, it was
demonstrated that after rapid, deep quench, the typical domain size
grows as a power-law function of time with the exponent of about 0.5
and then it gradually crosses over to logarithmically slow growth,
indicating strongly suppressed kinetics of phase
ordering~\cite{Kob11}.   In the previous study, it was found that
the distribution of pore sizes in highly porous systems is well
described by a scaling relation up to intermediate length scales,
while dense samples with only small isolated pores are characterized
by a Gaussian distribution of pore sizes~\cite{Makeev18}.  Moreover,
the analysis of local atomic density of solid domains revealed that,
with increasing porosity, the local density distribution function
develops a broad plateau and a strong peak in the vicinity of zero
density, which is characteristic of systems with large
surface-to-volume ratios~\cite{Makeev18}.  Nevertheless, the local
density of solid domains away from solid-void interfaces is rather
insensitive to porosity and its average value is only slightly below
the density of a homogeneous, pore-free glass phase~\cite{Makeev18}.

\vskip 0.05in


The mechanical properties and structural transformations of porous
glasses at \textit{isochoric conditions} were recently studied in
the cases of shear, tension, and compression using molecular
dynamics simulations~\cite{Priezjev17,Priezjev18t,Priezjev18c}. It
was demonstrated that in the linear regime of deformation, the
dependence of shear modulus on the average glass density is well
described by a power-law function with the exponent of
2.41~\cite{Priezjev17}. Interestingly, the same exponent was found
to hold for the density-dependence of the elastic modulus during
tensile and compressive deformations, despite the finite value of
normal stress at zero strain that appears in porous samples formed
at constant volume~\cite{Priezjev18t,Priezjev18c}. In addition, it
was shown that the variation of shear modulus on porosity at low
average glass densities is in good agreement with the continuum
predictions based on the percolation
theory~\cite{Sahimi94,Priezjev17}. Upon further increasing strain,
the pore shapes become significantly deformed and adjacent pores
coalesce with each other, leading to formation of large voids,
necking and eventually breaking of the
material~\cite{Priezjev17,Priezjev18t}.  At high compressive strain,
the pore coalescence and void redistribution result in the formation
of nearly homogeneous solid domains, which provide enhanced
resistance to deformation~\cite{Priezjev18c}. However, the influence
of various deformation protocols (\textit{e.g.}, constant pressure
vs. constant volume deformation and stress vs. strain-controlled
loading) on mechanical and structural properties of porous glasses
remains not fully understood.

\vskip 0.05in

In this paper, the structural relaxation of porous glasses that are
initially prepared at constant volume, is studied in the $NPT$
ensemble using molecular dynamics simulations. It will be shown that
porous samples become significantly compressed under built-in
tensile stresses, which is reflected in the time evolution of the
average glass density and the shape of the pore size distribution
functions.  The slow aging process is also characterized by the
gradual decrease in the potential energy.  The mechanical properties
are probed at different waiting times by imposing tensile strain at
constant pressure. It is found that the functional form of the
elastic modulus as a function of the average glass density holds for
tensile loading at constant volume as well as at constant pressure
and different waiting times. The analysis of the atomic
configurations and pore size distributions helps to distinguish the
process of breaking of the porous samples at large strain for lower
average glass densities from the formation of a nearly homogeneous,
high-density amorphous material.

\vskip 0.05in

The paper is organized as follows. The details of molecular dynamics
simulations as well as the preparation and deformation protocols are
described in the next section. The analysis of the pore size
distributions, potential energy, density profiles, and mechanical
properties of porous glasses are presented in
Sec.\,\ref{sec:Results}. The brief summary is given in the last
section.

\section{MD simulation details}
\label{sec:MD_Model}


The molecular dynamics simulations described below were performed on
a model glass former represented by the Kob-Andersen (KA) binary
mixture~\cite{KobAnd95}.  Our system consists of $240\,000$ large
atoms of type $A$ and $60\,000$ small atoms of type $B$, thus making
the total number of atoms $N=300\,000$.   Upon cooling, the
crystallization of the KA mixture is suppressed because of the
non-additive Lennard-Jones (LJ) interactions~\cite{KobAnd95}. More
specifically, the interaction between any two atoms of types
$\alpha,\beta=A,B$ is defined by the LJ potential:
\begin{equation}
V_{\alpha\beta}(r)=4\,\varepsilon_{\alpha\beta}\,\Big[\Big(\frac{\sigma_{\alpha\beta}}{r}\Big)^{12}\!-
\Big(\frac{\sigma_{\alpha\beta}}{r}\Big)^{6}\,\Big],
\label{Eq:LJ_KA}
\end{equation}
with the following parametrization $\varepsilon_{AA}=1.0$,
$\varepsilon_{AB}=1.5$, $\varepsilon_{BB}=0.5$, $\sigma_{AB}=0.8$,
and $\sigma_{BB}=0.88$~\cite{KobAnd95}. The mass of each atom type
is the same, $m_{A}=m_{B}$.  To speed up computation, the LJ
potential is truncated at the cutoff radius
$r_{c,\,\alpha\beta}=2.5\,\sigma_{\alpha\beta}$. Throughout the
paper, the results are reported in the reduced units of length,
mass, energy, and time, \textit{i.e.}, $\sigma=\sigma_{AA}$,
$m=m_{A}$, $\varepsilon=\varepsilon_{AA}$, and
$\tau=\sigma\sqrt{m/\varepsilon}$, respectively.  The numerical
integration of the Newton's equations of motion was carried out
using the velocity-Verlet algorithm~\cite{Allen87,Lammps} with the
time step $\triangle t_{MD}=0.005\,\tau$.

\vskip 0.05in


The initial equilibration of the system in the periodic box of
constant volume and the subsequent thermal quench across the glass
transition is identical to the preparation procedure implemented in
the previous MD
studies~\cite{Kob11,Kob14,Priezjev17,Makeev18,Priezjev18t,Priezjev18c}.
Here, we briefly describe the numerical protocol, which results in
the formation of the porous glass.   The system was initially
equilibrated at constant volume and the temperature of
$1.5\,\varepsilon/k_B$ during $3\times 10^4\,\tau$.   The Boltzmann
constant is denoted by $k_B$.   This temperature is well above the
glass transition temperature of the KA model
$T_g\approx0.435\,\varepsilon/k_B$ at the atomic density
$\rho=\rho_{A}+\rho_{B}=1.2\,\sigma^{-3}$~\cite{KobAnd95}. Next, the
temperature was instantaneously reduced to the target value
$T=0.05\,\varepsilon/k_B$, and the system was allowed to evolve
freely at constant volume during the time interval of
$10^{4}\,\tau$. At the low temperature, the phase separation
kinetics leads to the formation of porous structure in the amorphous
solid~\cite{Kob11,Kob14,Priezjev17,Makeev18,Priezjev18t,Priezjev18c}.
In the present study, the simulation results were carried out only
for one realization of disorder due to computational limitations.

\vskip 0.05in


After the porous glass samples were prepared at constant volume, the
follow-up MD simulations were performed in the isothermal-isobaric
($NPT$) ensemble, where $P=0$ and $T=0.05\,\varepsilon/k_B$.  The
temperature $T=0.05\,\varepsilon /k_{B}$ was regulated via the
Nos\'{e}-Hoover thermostat~\cite{Lammps}. As discussed in detail in
the next section, the relaxation process at constant pressure
results in the gradual increase of the average glass density even in
the absence of external deformation.  The periodic boundary
conditions were applied in all three dimensions, which were allowed
to change independently during densification of the porous samples.
Next, the tensile deformation was imposed along the $\hat{x}$
direction with the strain rate
$\dot{\varepsilon}_{xx}=10^{-4}\,\tau^{-1}$, while the pressure in
the $\hat{y}$ and $\hat{z}$ directions was maintained at zero. The
maximum strain was varied from 100\% to 400\% depending on the
average glass density.  During the production runs, the pressure
components, potential energy, system dimensions as well as atomic
configurations were saved for the postprocessing analysis of the
porous structure and mechanical properties.

\section{Results}
\label{sec:Results}

\begin{figure}[t]
\includegraphics[width=15.cm,angle=0]{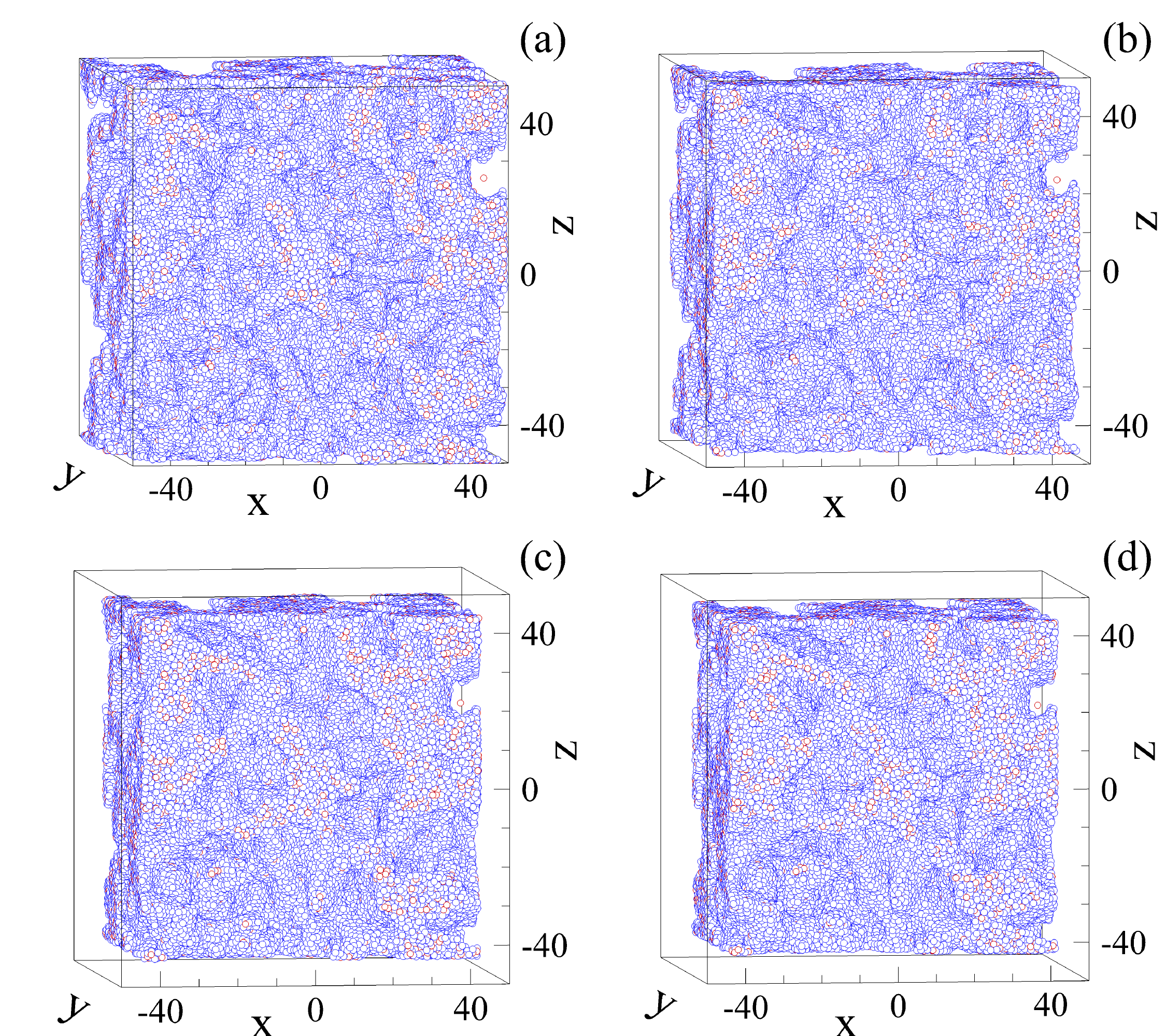}
\caption{(Color online) A series of snapshots of the porous glass at
zero pressure after time intervals (a) $0\,\tau$, (b) $500\,\tau$,
(c) $50\,000\,\tau$, and (d) $250\,000\,\tau$.  The average glass
densities are (a) $\rho\sigma^{3}=0.30$, (b) $\rho\sigma^{3}=0.37$,
(c) $\rho\sigma^{3}=0.49$, and (d) $\rho\sigma^{3}=0.52$. The
temperature is $T=0.05\,\varepsilon/k_B$ and the total number of
atoms is $N=300\,000$. Atoms of types $A$ and $B$ are denoted by
blue and red spheres.  Note that the size of atoms is reduced for
clarity. }
\label{fig:snapshot_system}
\end{figure}

As discussed in the previous section, the formation of porous glassy
systems occurs in the process of concurrent phase separation and
material solidification after sufficiently deep thermal quench from
a liquid state~\cite{Kob11,Kob14}. The kinetics of the process at
low temperature and constant volume becomes logarithmically slow and
the particle diffusion in the glass phase is strongly
suppressed~\cite{Kob11,Kob14}.   An example of the porous structure
in the amorphous solid formed at constant volume after the long time
interval of $10^{4}\,\tau$ is shown in
Fig.\,\ref{fig:snapshot_system}\,(a) for the average glass density
$\rho\sigma^{3}=0.3$.    Notice the complex interconnected pore
network embedded into a dense glass phase.  Thus formed porous
samples in a wide range of average glass densities, $0.2 \leqslant
\rho\sigma^3 \leqslant 1.0$, were used as starting configurations
for the MD simulations performed in the $NPT$ ensemble and described
in the present study.

\vskip 0.05in


It should be particularly emphasized that the porous glass formation
at constant volume and low temperature occurs at a negative
pressure, which depends on the target temperature and the average
glass density~\cite{Makeev18}. Moreover, it was recently deduced
from extensive molecular dynamics simulations that the data in a
wide range of parameters can be collapsed onto a scaling relation of
the form  $P/T\sim \rho^{\alpha}$ with the exponent $\alpha\approx
2.5$~\cite{Makeev18}.   Thus, at the temperature
$T=0.05\,\varepsilon/k_B$ considered in the present study, the
average values of pressure vary from
$P\approx-0.73\,\varepsilon/\sigma^3$ for $\rho\sigma^{3}=0.9$ to
$P\approx-0.01\,\varepsilon/\sigma^3$ for $\rho\sigma^{3}=0.2$. In
other words, the glassy porous systems can be viewed as being under
tension in a confined geometry.   It is, therefore, expected that
when the constraint of constant volume is relaxed and the
simulations are instead performed at constant pressure, the porous
samples will undergo isotropic compression, which is driven by the
build-in tensile stresses.   The evolution of atomic configurations
at zero pressure and temperature $T=0.05\,\varepsilon/k_B$ is shown
in Fig.\,\ref{fig:snapshot_system} for the indicated waiting times.
It can be clearly observed that the solid phase and the porous
network become gradually compressed and the average glass density
increases over time.

\vskip 0.05in

\begin{figure}[t]
\includegraphics[width=12.cm,angle=0]{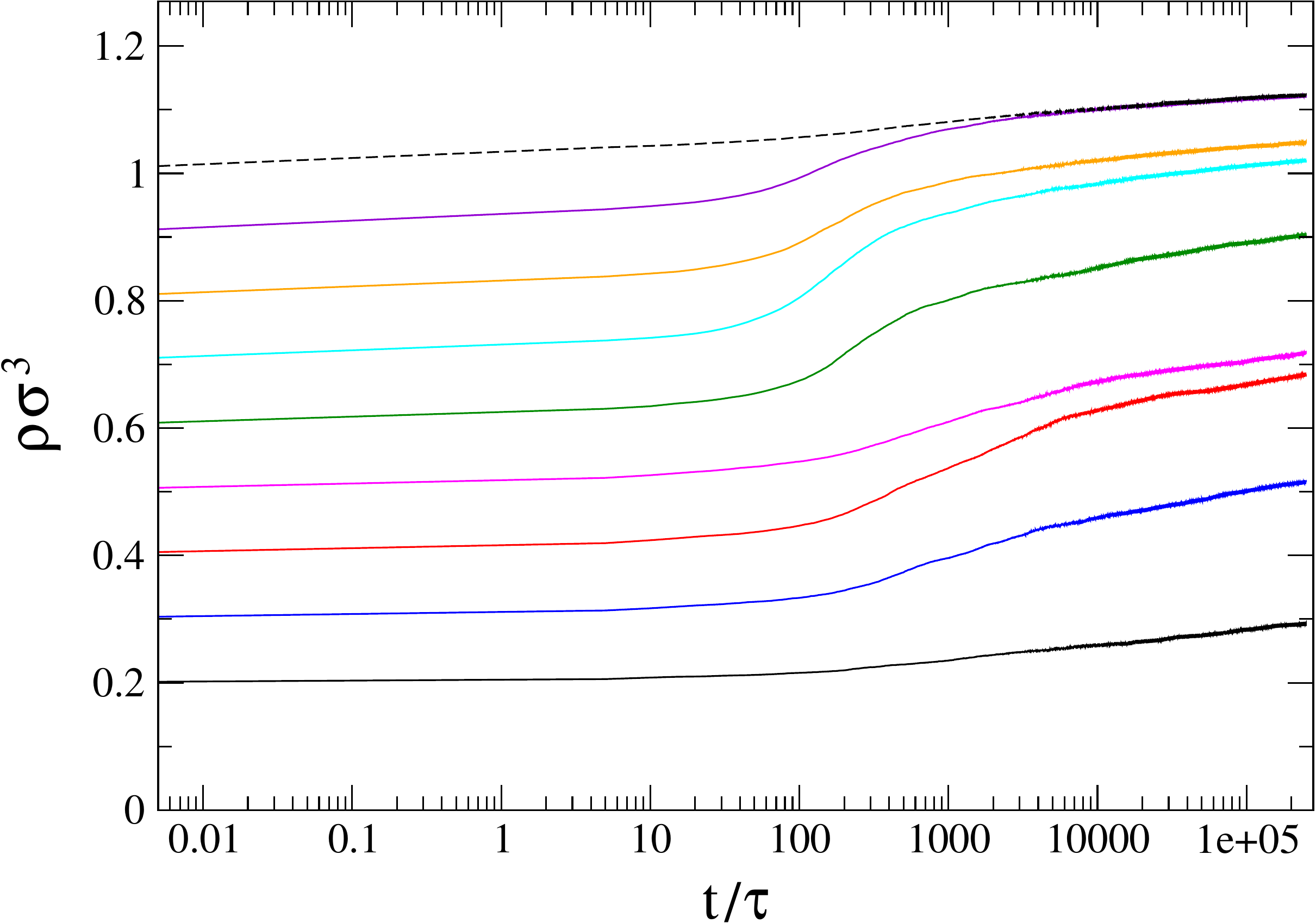}
\caption{(Color online) The average glass densities as a function of
waiting time during the aging process at zero pressure and
temperature $T=0.05\,\varepsilon/k_B$.  The average density of
porous samples at $t=0$ is $\rho\sigma^{3}=0.2$, $0.3$, $0.4$,
$0.5$, $0.6$, $0.7$, $0.8$, $0.9$, and $1.0$ (from bottom to top).}
\label{fig:density_age_time}
\end{figure}
\begin{figure}[t]
\includegraphics[width=12.cm,angle=0]{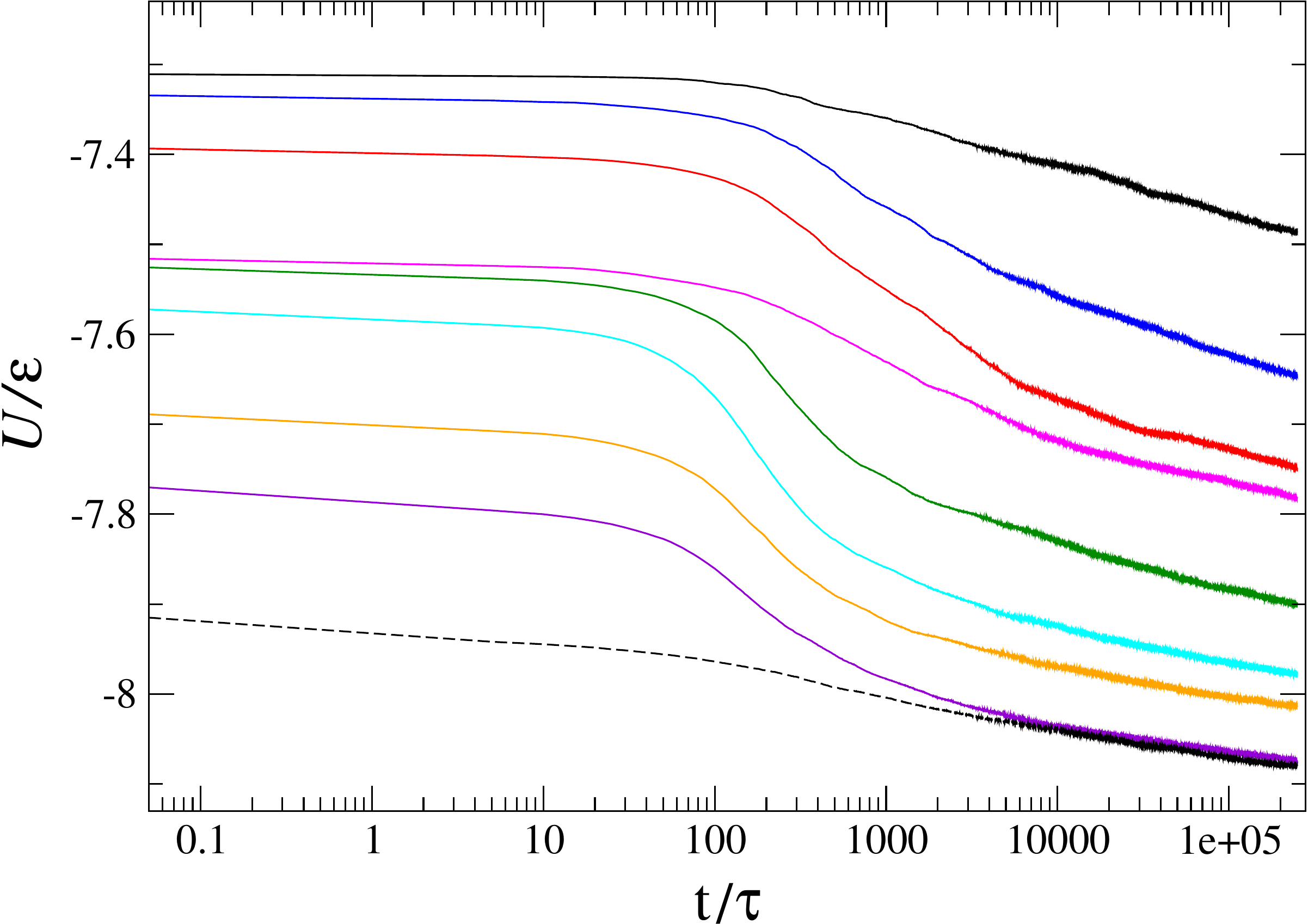}
\caption{(Color online) The time dependence of the potential energy
per atom for nine samples at constant pressure $P=0$ and
$T=0.05\,\varepsilon/k_B$. The average glass density increases from
top to bottom. The same samples and color codes as in
Fig.\,\ref{fig:density_age_time}. }
\label{fig:poten_age_time}
\end{figure}

A more quantitative analysis of the relaxation process under
internal stresses at constant pressure, $P=0$, can be performed by
considering the time dependence of the average glass density and the
potential energy. First, the variation of the average glass density
as a function of the waiting time is plotted in
Fig.\,\ref{fig:density_age_time}.   It can be seen that the most
rapid densification occurs after the time interval of about
$100\,\tau$, which is followed by a slow aging process up to
$2.5\times10^5\,\tau$.  As evident, the relative increase in the
average glass density during the time interval of
$2.5\times10^5\,\tau$ is more pronounced in lower density samples.
Note also that the average glass density of the two dense samples
(two upper curves in Fig.\,\ref{fig:density_age_time}) becomes
nearly the same after about $10^4\,\tau$.    We comment that these
dense samples are not pore-free even after the time interval of
$2.5\times10^5\,\tau$ as their average density is still below the
density of the homogeneous glass phase of about $1.21\,\sigma^{-3}$
at zero pressure~\cite{NVP18strload}.   Second, the time dependence
of the potential energy per atom for the same porous samples is
reported in Fig.\,\ref{fig:poten_age_time}. Similar to the case of
density, the potential energy starts to decrease rapidly after about
$100\,\tau$ followed by a gradual decay during the next two decades,
which suggests that the relaxation dynamics becomes progressively
more slow.  The lowest potential energy is also attained in two
dense samples at $t\gtrsim 10^4\,\tau$.


\vskip 0.05in


The analysis of pore structure and porosity evolution in the systems
under consideration is based on calculation of the pore size
distribution (PSD) functions. To this end, the open-source Zeo++
software was utilized~\cite{Haranczyk12c,Haranczyk12,Haranczyk17}.
The pore sizes were computed using algorithms implemented in
Zeo++~\cite{Haranczyk12c,Haranczyk12,Haranczyk17}. The approach
involves the Voronoi decomposition of the total volume of system
into Voronoi cells, associated with each individual atom in the
system. The implementation derives from a modification to the VORO++
software library, developed in Ref.\,\cite{Rycroft09}. This
computational tool provides Voronoi network and all the parameters
necessary for analysis of the geometrical characteristics of void
space in porous material systems. The Voronoi decomposition is
performed such that the space surrounding atoms is divided into
polyhedral cells and each face of the polyhedral is a plane
equidistant from the two points sharing the face. Edges of the cells
correspond to lines equidistant to neighboring points, whereas
vertices are equidistant from neighboring points. The Voronoi
network, built thereby, maps the void space in porous systems.
Further, the computed Voronoi network can be utilized to obtain
quantitative information on the largest pore in the systems and the
largest pore that can traverse through it, topological properties of
pores and channels, and other geometric quantities. It should be
noted that Zeo++ allows one to obtain information for model systems
with periodic structural units, comprised of multiple types of
atoms.  For a given radius of the probe, one can identify the probe
accessible regions of the void-space network using a graph
propagation algorithm. In practice, a variation of the Dijkstra
shortest path algorithm is used~\cite{Dijkstra59}. In this work, the
analysis was used to identify the topological properties of channels
in the porous systems. The surface areas of the pores and their
volumes are computed using a Monte Carlo sampling. The number of
samples per atom used in this works is fixed at 50000. The probe
radius used for calculations is $0.3\,\sigma$. The value of the
probe radius was varied in a wide range and no significant
variations in the results were observed for values less than
$1.0\,\sigma$.

\vskip 0.05in

\begin{figure}[t]
\includegraphics[width=12.cm,angle=0]{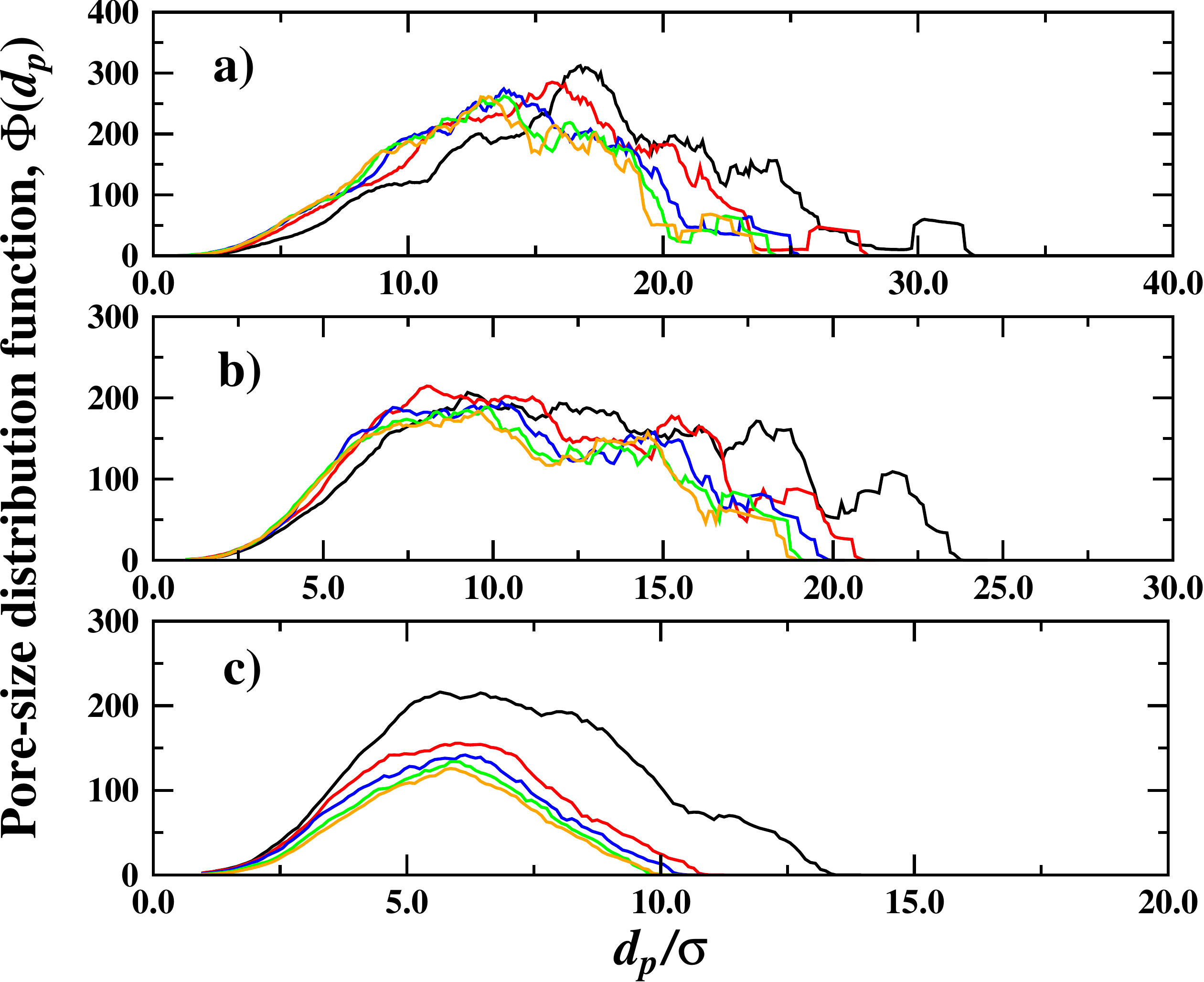}
\caption{(Color online) The pore size distributions in quiescent
samples during the structural relaxation process at constant
pressure. The average glass densities at $t=0$ are (a)
$\rho\sigma^{3}=0.3$, (b) $\rho\sigma^{3}=0.5$, and (c)
$\rho\sigma^{3}=0.8$. The distribution functions for different
waiting times are indicated by black curves ($t=0$), red curves
($t=5\times 10^2\,\tau$), blue curves ($t=5\times 10^3\,\tau$),
green curves ($t=5\times 10^4\,\tau$), and orange curves
($t=2.5\times 10^5\,\tau$). Note that scales are different in three
panels. }
\label{fig:pore_size_dist_age}
\end{figure}

The pore size distribution functions during the relaxation process
at constant pressure are presented in
Fig.\,\ref{fig:pore_size_dist_age} for the indicated waiting times.
These data were computed in three porous samples after the constant
pressure condition was applied in the absence of external loading.
It can be clearly observed that with increasing average glass
density, the pore size distributions at $t=0$ become more narrow.
This behavior is consistent with the results of the previous MD
studies, where simulations were performed at constant
volume~\cite{Priezjev17,Makeev18,Priezjev18t,Priezjev18c}. The
distribution functions presented in
Fig.\,\ref{fig:pore_size_dist_age}\,(a) correspond to atomic
configurations shown in Fig.\,\ref{fig:snapshot_system} as well as
blue curves in Figs.\,\ref{fig:density_age_time} and
\ref{fig:poten_age_time}.   It can be seen that, as the pore network
becomes more compact, the shape of distributions are skewed toward
smaller length scales.  In agreement with the slow variation of the
average glass density and potential energy at large waiting times,
the PSDs at $t=5\times 10^4\,\tau$ and $2.5\times 10^5\,\tau$ are
nearly the same, apart from statistical fluctuations.  The case of
higher density porous glass shown in
Fig.\,\ref{fig:pore_size_dist_age}\,(b) appears to be qualitatively
similar, although the average pore size is smaller.  In contrast,
the pore size distributions in the case shown in
Fig.\,\ref{fig:pore_size_dist_age}\,(c) evolve to the symmetric,
bell-shaped curve at large times, which is representative of
low-porosity systems with a collection of isolated
pores~\cite{Makeev18}.


\vskip 0.05in

\begin{figure}[t]
\includegraphics[width=12.cm,angle=0]{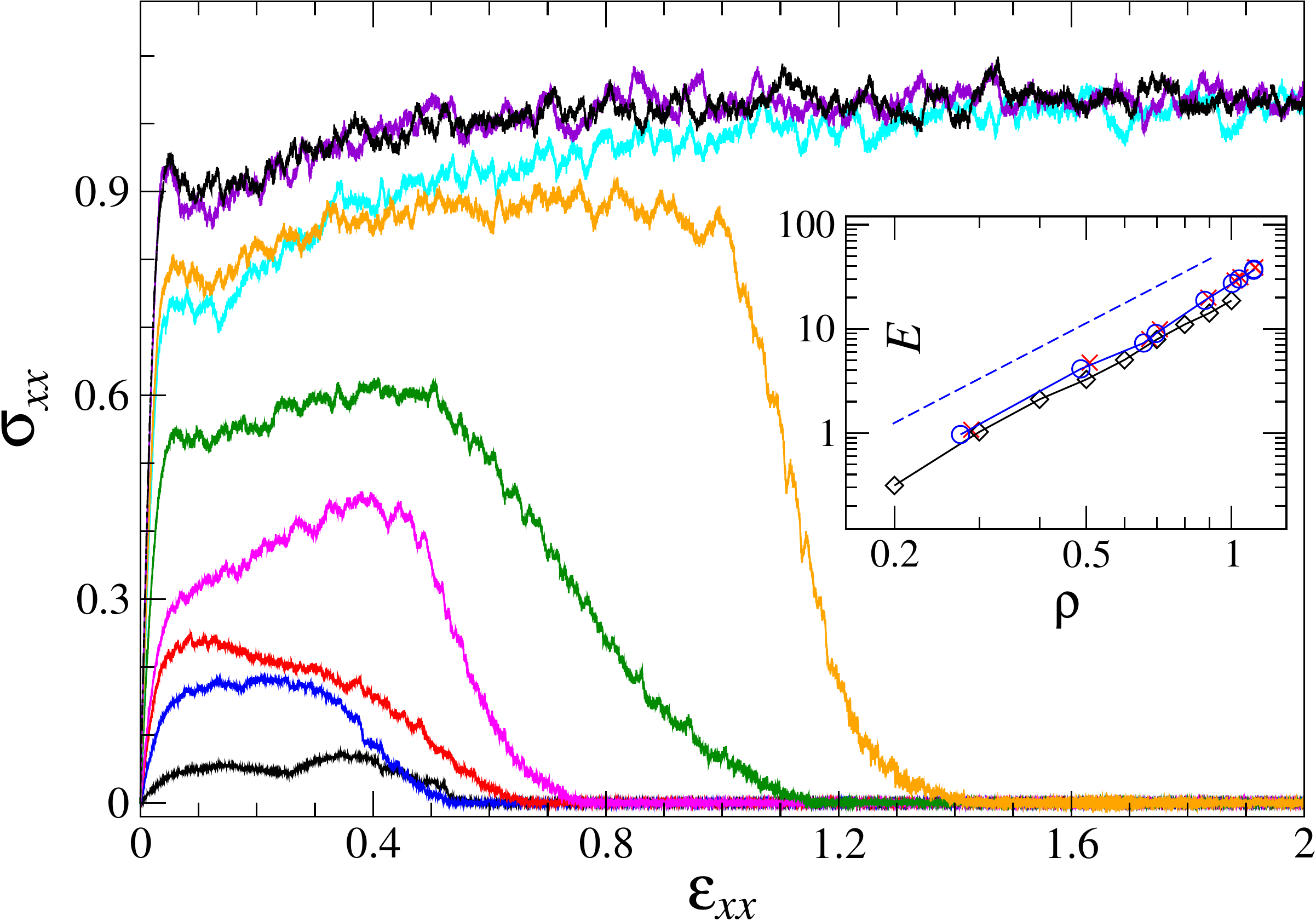}
\caption{(Color online) The variation of stress $\sigma_{xx}$ (in
units of $\varepsilon\sigma^{-3}$) as a function of strain during
tensile loading at zero pressure with the strain rate
$\dot{\varepsilon}_{xx}=10^{-4}\,\tau^{-1}$. The samples with
different average glass densities were strained after the waiting
time of $5\times 10^4\,\tau$ (see text for details).  The inset
shows the elastic modulus $E$ (in units of $\varepsilon\sigma^{-3}$)
as a function of the average glass density $\rho\sigma^{3}$. The
data in the inset were computed after the waiting time of
$5\times10^4\,\tau$ (blue circles) and after $15\times10^4\,\tau$
(red crosses).  For reference, the data for the elastic modulus at
constant volume (taken from Ref.\,\cite{Priezjev18t}) are also shown
by black diamonds. The straight dashed line indicates the slope of
$2.41$. }
\label{fig:stress_strain}
\end{figure}

We next discuss the mechanical response of the porous glass to
tensile loading at constant pressure $P=0$ and temperature
$T=0.05\,\varepsilon/k_B$.  In Fig.\,\ref{fig:stress_strain}, the
stress-strain curves are plotted for nine samples with different
average glass densities after the waiting time of $5\times
10^4\,\tau$.  In can be seen that both the slope of the linear
region and the yield stress become larger with increasing average
glass density.  We comment that in contrast to a finite stress at
zero strain in porous glasses prepared and strained at constant
volume~\cite{Priezjev18t,Priezjev18c}, the stress is initially zero
in all samples shown in Fig.\,\ref{fig:stress_strain}. Furthermore,
at larger strain, $\varepsilon_{xx} \gtrsim 0.1$ in
Fig.\,\ref{fig:stress_strain}, the stress curves exhibit a shallow
maximum and then decay to zero, indicating breakup of the amorphous
material into separate domains (discussed below). Notice that the
failure of denser samples occurs at higher values of strain.
However, in the three cases of higher density samples, the stress
appears to saturate to a broad plateau, which extends up to
$\varepsilon_{xx} = 2.0$.   The stress amplitude at the plateau
level is determined by the deformation of a nearly homogeneous glass
phase, where almost all pores got transformed and ejected from the
amorphous material.

\vskip 0.05in


The elastic modulus, $E$, computed in the linear regime of
deformation after the waiting time of $5\times 10^4\,\tau$ is shown
in the inset to Fig.\,\ref{fig:stress_strain} as a function of the
average glass density.  As is evident, the elastic modulus follows a
power-law dependence $E \sim \rho^{2.41}$, which is indicated by the
straight dashed line.  Interestingly, the same exponent was reported
in our previous studies of tension, compression, and shear of porous
glasses at constant
volume~\cite{Priezjev17,Priezjev18t,Priezjev18c}. For reference, the
data for tension of porous glasses at constant
volume~\cite{Priezjev18t} are also plotted in the inset to
Fig.\,\ref{fig:stress_strain}.  As shown in
Fig.\,\ref{fig:density_age_time}, the average glass density in all
samples gradually increases over time even in the absence of
external deformation, and, therefore, it is expected that the
elastic modulus will be larger for more aged samples.  Indeed, the
data collected after longer waiting time of $15\times 10^4\,\tau$
reveal that although the elastic modulus and the average glass
density slightly increase in each sample, the data still follow the
same power-law dependence as for the less aged glasses (see inset in
Fig.\,\ref{fig:stress_strain}).   These results demonstrate that the
functional form of the density-dependent elastic modulus of porous
glasses is the same for both deformation protocols, \textit{i.e.},
tension at constant volume and constant pressure.

\vskip 0.05in

\begin{figure}[t]
\includegraphics[width=15.cm,angle=0]{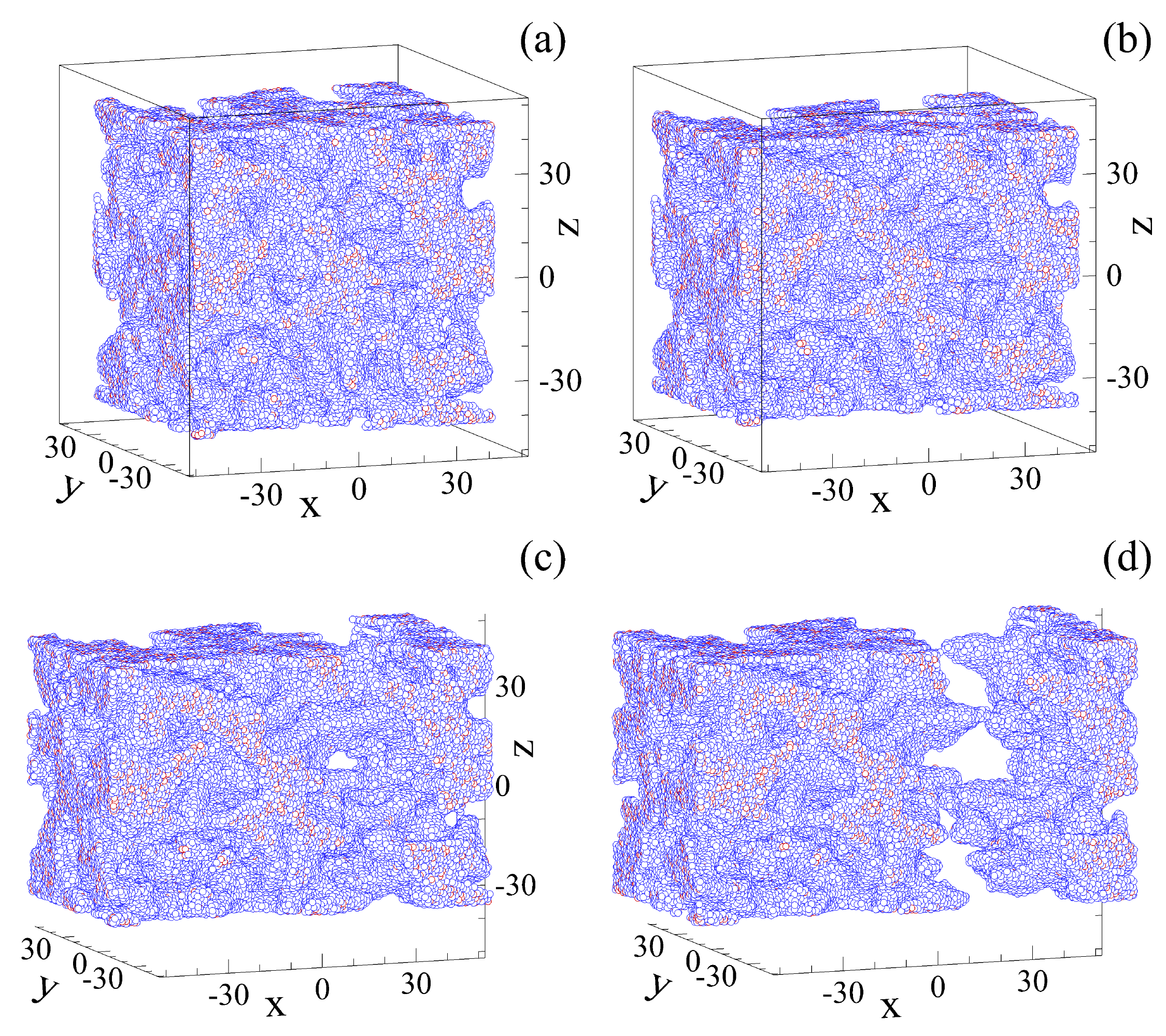}
\caption{(Color online) Atomic configurations of the strained porous
glass at zero pressure and temperature $T=0.05\,\varepsilon/k_B$.
The values of strain and the average glass density are (a)
$\varepsilon_{xx}=0.05$ and $\rho\sigma^{3}=0.48$, (b)
$\varepsilon_{xx}=0.20$ and $\rho\sigma^{3}=0.49$, (c)
$\varepsilon_{xx}=0.40$ and $\rho\sigma^{3}=0.47$, and (d)
$\varepsilon_{xx}=0.60$ and $\rho\sigma^{3}=0.41$. The strain rate
is $\dot{\varepsilon}_{xx}=10^{-4}\,\tau^{-1}$. The stress-strain
curve for this sample is marked by blue color in
Fig.\,\ref{fig:stress_strain}.}
\label{fig:snapshot_strain_rho03}
\end{figure}
\begin{figure}[t]
\includegraphics[width=15.cm,angle=0]{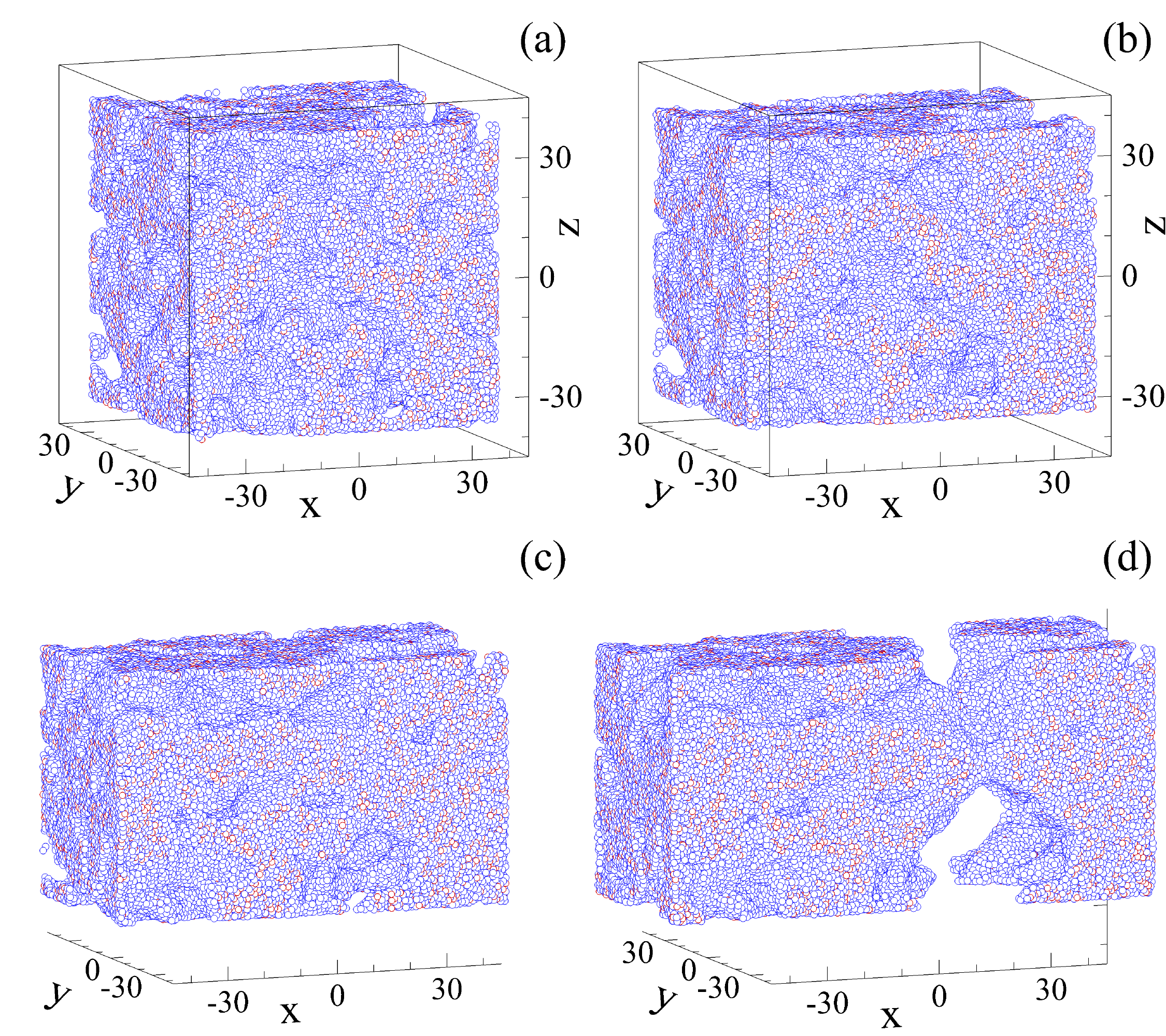}
\caption{(Color online) Instantaneous snapshots for the values of
strain and the average glass density (a) $\varepsilon_{xx}=0.05$ and
$\rho\sigma^{3}=0.69$, (b) $\varepsilon_{xx}=0.20$ and
$\rho\sigma^{3}=0.73$, (c) $\varepsilon_{xx}=0.40$ and
$\rho\sigma^{3}=0.80$, and (d) $\varepsilon_{xx}=0.70$ and
$\rho\sigma^{3}=0.70$. The stress is denoted by the magenta curve in
Fig.\,\ref{fig:stress_strain}. }
\label{fig:snapshot_strain_rho05}
\end{figure}
\begin{figure}[t]
\includegraphics[width=15.cm,angle=0]{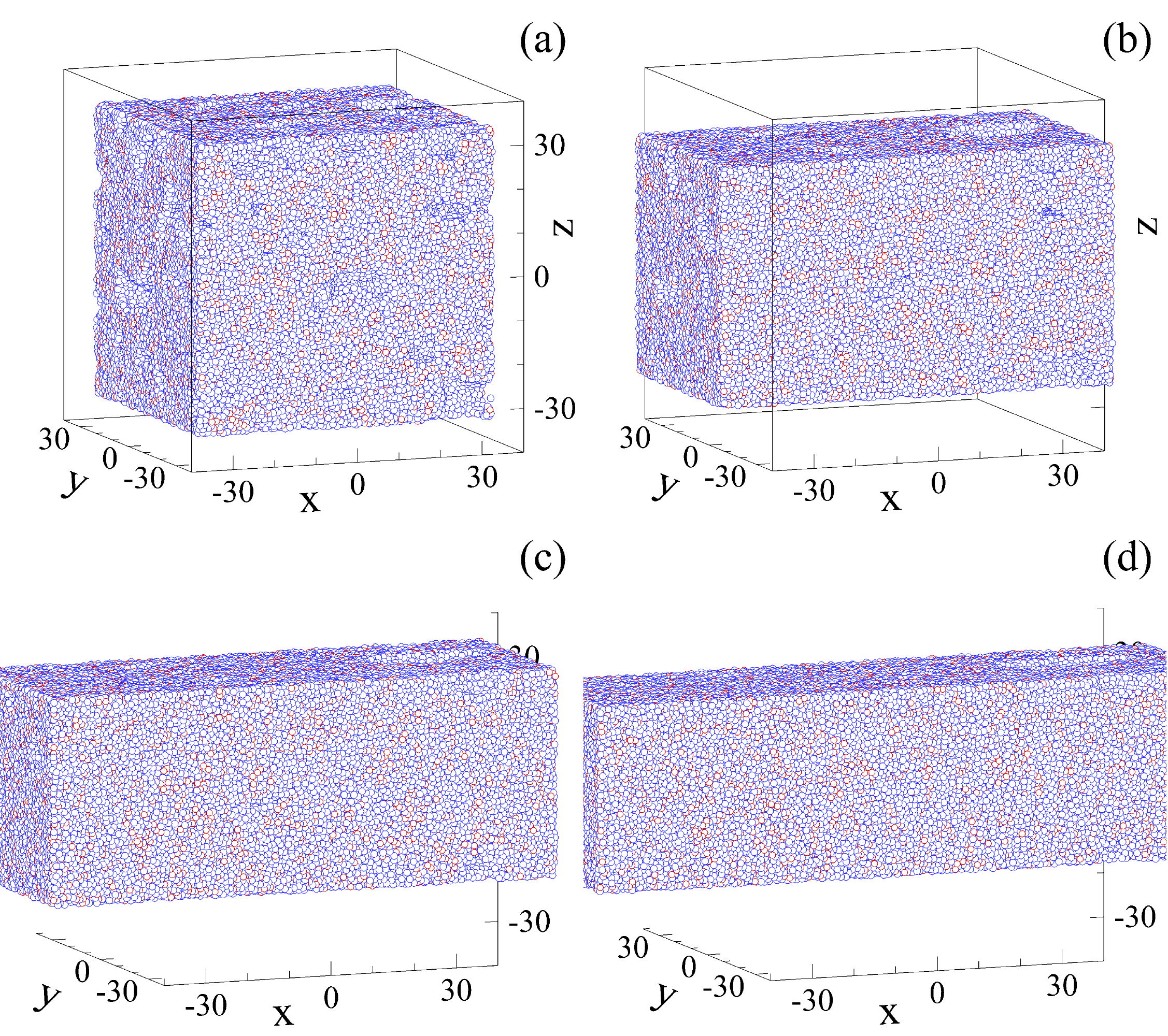}
\caption{(Color online) Selected snapshots of the porous glass for
strains and average glass densities (a) $\varepsilon_{xx}=0.05$ and
$\rho\sigma^{3}=1.0$, (b) $\varepsilon_{xx}=0.40$ and
$\rho\sigma^{3}=1.11$, (c) $\varepsilon_{xx}=0.80$ and
$\rho\sigma^{3}=1.16$, and (d) $\varepsilon_{xx}=1.60$ and
$\rho\sigma^{3}=1.18$. The stress dependence on strain is indicated
by the cyan curve in Fig.\,\ref{fig:stress_strain}. }
\label{fig:snapshot_strain_rho07}
\end{figure}

The representative snapshots of strained porous glasses after the
waiting time of $5\times 10^4\,\tau$ are presented in
Figs.\,\ref{fig:snapshot_strain_rho03},
\ref{fig:snapshot_strain_rho05}, and \ref{fig:snapshot_strain_rho07}
for different average glass densities.  The atomic configurations
for three samples are shown for the indicated values of strain along
the stress-strain curves reported in Fig.\,\ref{fig:stress_strain}.
It can be clearly seen that the pore configurations in less dense
samples (shown in Figs.\,\ref{fig:snapshot_strain_rho03} and
\ref{fig:snapshot_strain_rho05}) become significantly distorted,
leading to necking and eventually breaking of the material at large
strain.   These results are qualitatively similar to the cases of
tension at constant volume reported in the previous MD
study~\cite{Priezjev18t}.    Note, however, that the average glass
density increases slightly during tension at constant pressure,
except at large strain in
Figs.\,\ref{fig:snapshot_strain_rho03}\,(d) and
\ref{fig:snapshot_strain_rho05}\,(d) where the formation of
system-spanning voids contributes to the apparent decrease in the
average glass density.   By contrast, the deformation of the dense
porous sample proceeds via the formation of increasingly dense and
homogeneous glass phase (see Fig.\,\ref{fig:snapshot_strain_rho07}).
Notice that some isolated pores become significantly elongated in
highly strained samples.   We also checked that necking is not
formed in the dense porous glass during tension at constant pressure
up to $\varepsilon_{xx} \leqslant 4.0$ (not shown).

\vskip 0.05in

\begin{figure}[t]
\includegraphics[width=12.cm,angle=0]{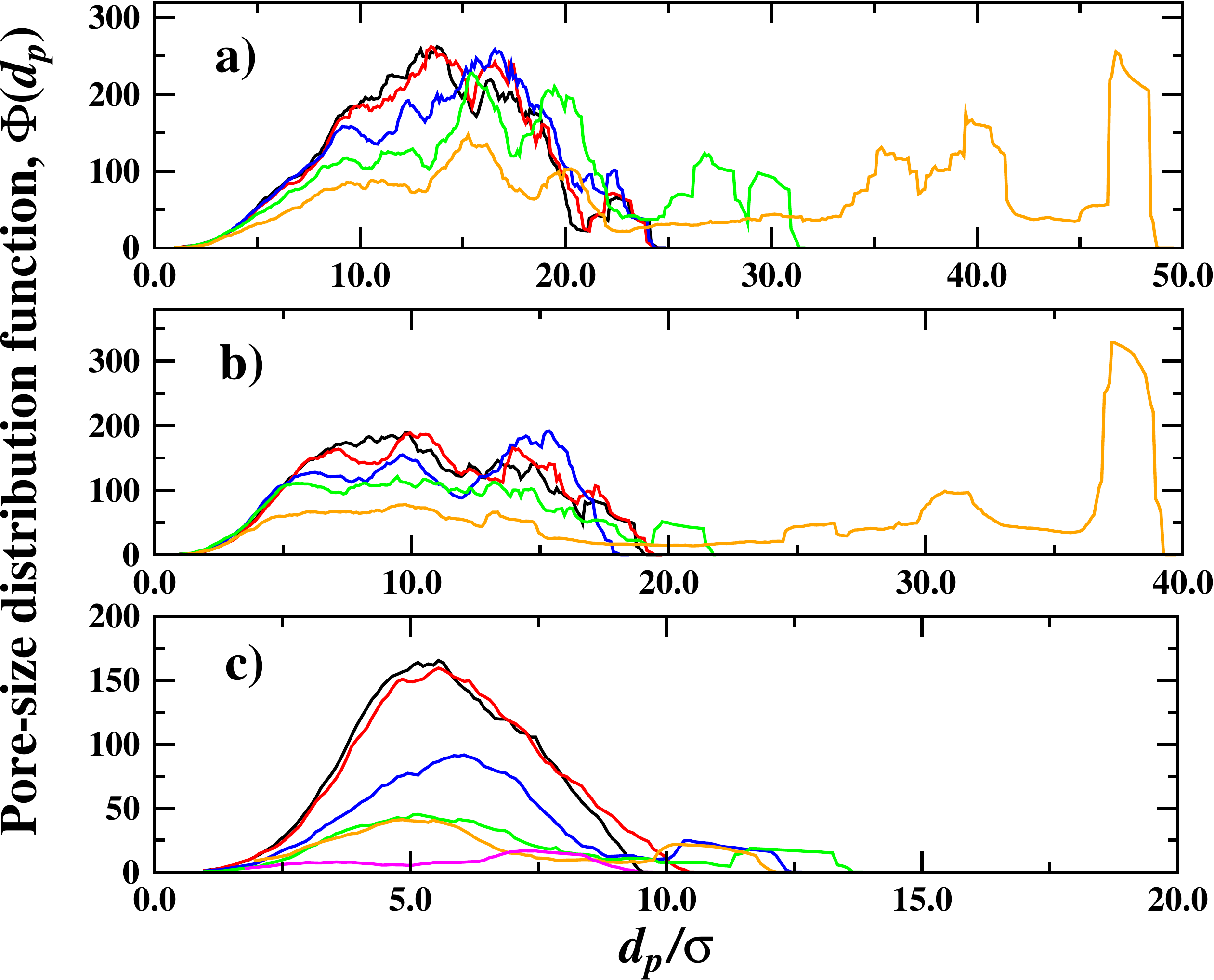}
\caption{(Color online) The pore size distribution functions for
three samples strained after the waiting time of $5\times
10^4\,\tau$.  The colorcode for different curves is black
($\varepsilon_{xx}=0.0$), red ($\varepsilon_{xx}=0.05$), blue
($\varepsilon_{xx}=0.20$), green ($\varepsilon_{xx}=0.40$), orange
($\varepsilon_{xx}=0.80$), and magenta ($\varepsilon_{xx}=1.60$).
The atomic configurations shown in
Figs.\,\ref{fig:snapshot_strain_rho03},
\ref{fig:snapshot_strain_rho05}, and \ref{fig:snapshot_strain_rho07}
are described by the pore size distributions presented in the panels
(a), (b), and (c), respectively. }
\label{fig:pore_size_dist_strain}
\end{figure}

The evolution of pore size distribution functions during tensile
loading after the waiting time of $5\times 10^4\,\tau$ are shown in
Fig.\,\ref{fig:pore_size_dist_strain} for the indicated values of
strain.  Note that the distributions plotted in the panels (a), (b),
and (c) in Fig.\,\ref{fig:pore_size_dist_strain} were computed using
atomic configurations presented in
Figs.\,\ref{fig:snapshot_strain_rho03},
\ref{fig:snapshot_strain_rho05}, and
\ref{fig:snapshot_strain_rho07}, respectively.  It can be seen in
Fig.\,\ref{fig:pore_size_dist_strain}\,(a-b) that lower density
porous samples show the same trend; namely, the pore size
distributions are initially narrow and peaked at $10 \lesssim
d_p/\sigma\lesssim 15$. With increasing strain, the PSDs become more
broad and a distinct peak develops at large length scales, which
corresponds to the formation of a large void associated with
material failure (see Figs.\,\ref{fig:snapshot_strain_rho03} and
\ref{fig:snapshot_strain_rho05}). These results are similar to the
breaking process during tensile loading at constant volume reported
in our previous study~\cite{Priezjev18t}.  By sharp contrast, the
pore size distributions for the higher density sample in
Fig.\,\ref{fig:pore_size_dist_strain}\,(c) remain confined to small
length scales and the amplitude of PSDs becomes significantly
reduced at high strain. This behavior correlates well with visual
observation of strained samples in
Fig.\,\ref{fig:snapshot_strain_rho07}, where only a few isolated,
highly deformed pores remain at high strain, thus forming a nearly
homogeneous glassy material.

\vskip 0.05in


As shown above, a significant rearrangement of the material takes
place during the transition to zero-pressure state as well as in the
process of mechanical loading. Therefore, it is important to
quantify the temporal evolution of density variations in the
systems. In our previous
studies~\cite{Priezjev17,Priezjev18t,Priezjev18c}, we analyzed
temporal evolution of the density profiles in model porous glasses
undergoing mechanical loading at constant volume. In what follows,
we apply a similar analysis to the atomic systems under
consideration. The methodology of analysis has been previously
described in detail in
Refs.\,\cite{Priezjev17,Priezjev18t,Priezjev18c}.  A brief summary
is as follows. We numerically compute spatially-resolved density
profiles along the direction of externally applied strain. The
quantity of interest is denoted by $\langle \rho \rangle_{s}(x)$ and
defined as the number of atoms of either type, located in a bin with
thickness $b\approx \sigma$ along the loading direction, and divided
by the volume of the bin, $b\,L_yL_z$, where $L_y$ and $L_z$ are the
box sizes in the two Cartesian directions perpendicular to the
loading direction.

\vskip 0.05in

\begin{figure}[t]
\includegraphics[width=12.cm,angle=0]{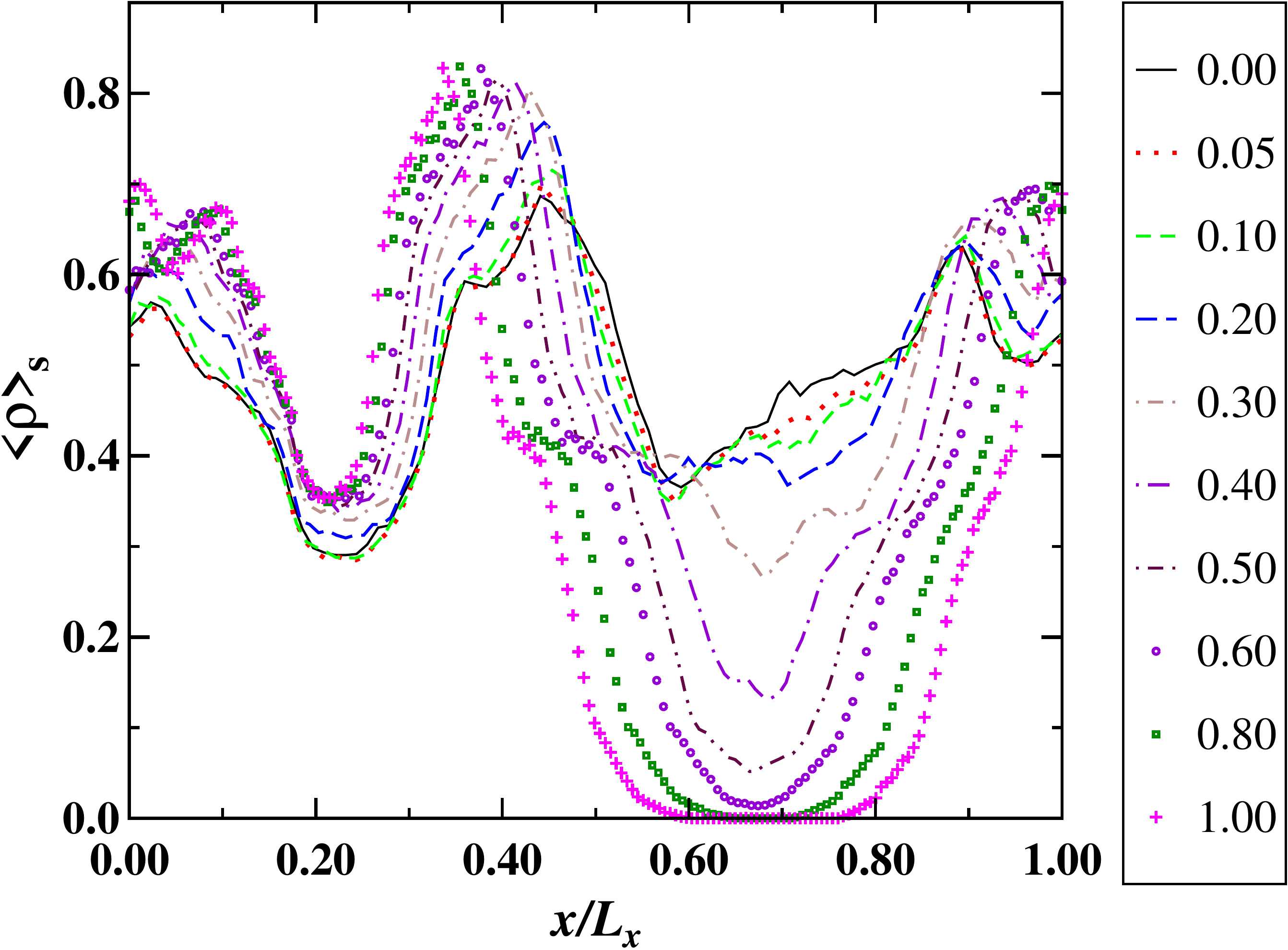}
\caption{(Color online) The averaged density profiles $\langle \rho
\rangle_{s}(x)$ (in units of $\sigma^{-3}$) computed along the
$x$-axis (corresponding to the loading direction), are plotted for
different values of the applied strain, as shown in the figure
legend. The system snapshots are presented in
Fig.\,\ref{fig:snapshot_strain_rho03} for four values of strain. }
\label{fig:den_prof_rho03}
\end{figure}
\begin{figure}[t]
\includegraphics[width=12.cm,angle=0]{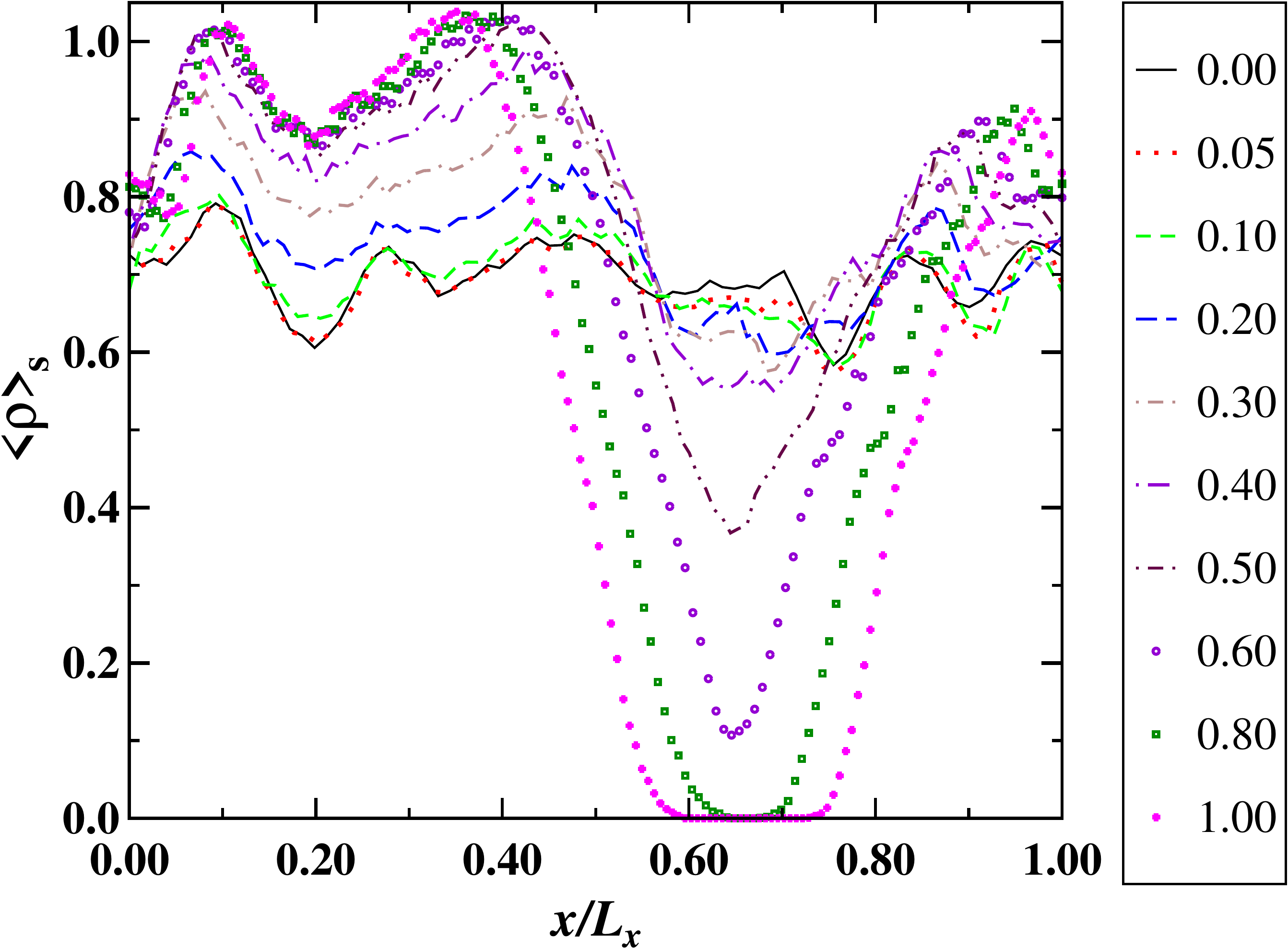}
\caption{(Color online) The local density profiles $\langle \rho
\rangle_{s}(x)$ (in units of $\sigma^{-3}$) along the loading
direction are presented for the indicated values of strain. Atomic
configurations for this sample are shown in
Fig.\,\ref{fig:snapshot_strain_rho05} for selected values of strain.
}
\label{fig:den_prof_rho05}
\end{figure}
\begin{figure}[t]
\includegraphics[width=12.cm,angle=0]{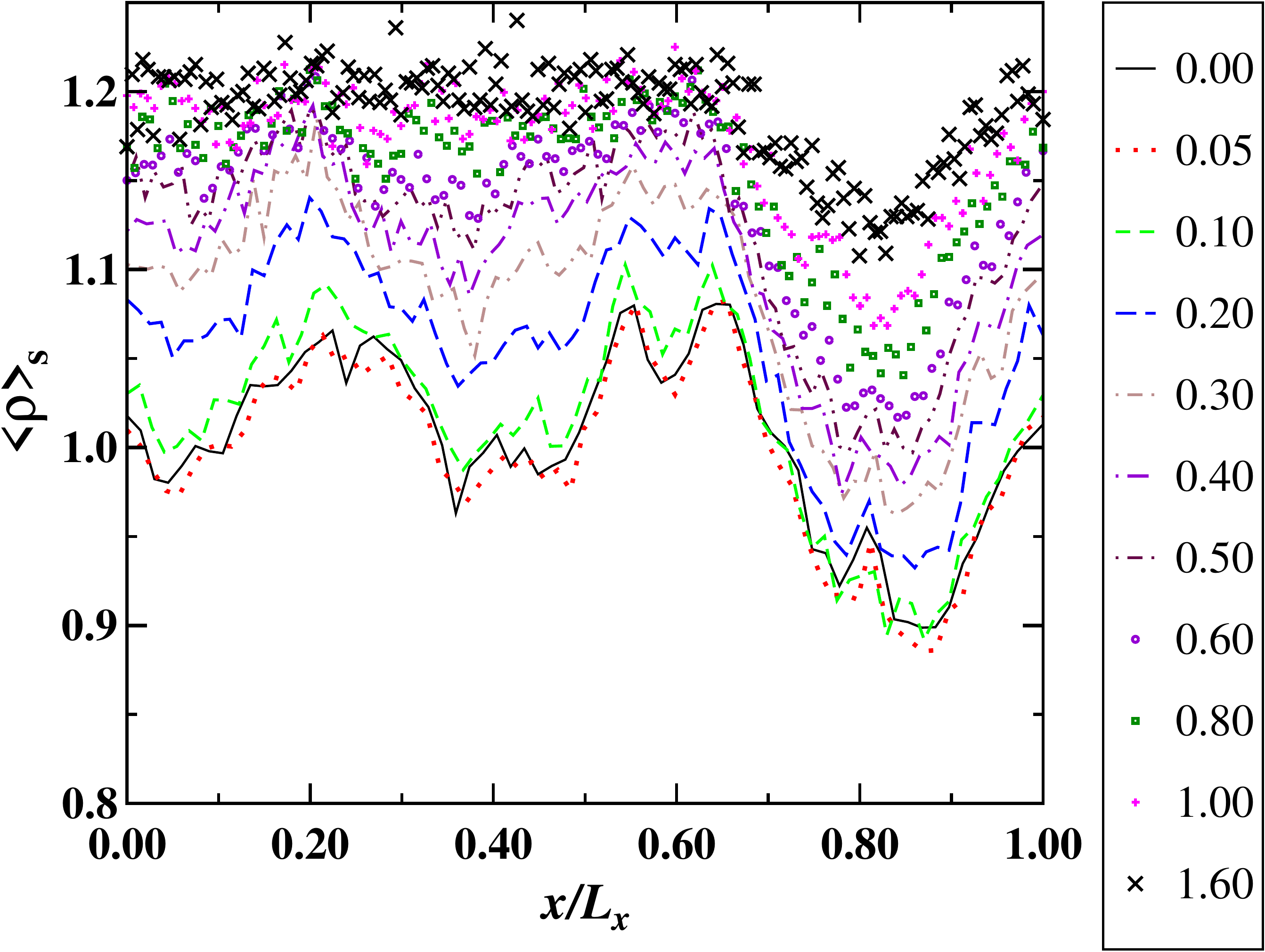}
\caption{(Color online) The density profiles $\langle \rho
\rangle_{s}(x)$ (in units of $\sigma^{-3}$) are plotted for the
values of strain listed in the legend. The corresponding atomic
configurations are displayed in
Fig.\,\ref{fig:snapshot_strain_rho07} for selected values of strain.
}
\label{fig:den_prof_rho07}
\end{figure}

Similar to the previous cases of tensile~\cite{Priezjev18t} and
compressive~\cite{Priezjev18c} loadings at constant volume, the
analysis of spatially-resolved density profiles reveals a number of
notable features pertained to density rearrangement during tension
at constant pressure. In particular, differences observed in systems
with low and high average densities are worth noting. The simulation
results are reported for three samples in
Figs.\,\ref{fig:den_prof_rho03}, \ref{fig:den_prof_rho05}, and
\ref{fig:den_prof_rho07}. At relatively low glass densities, shown
in Figs.\,\ref{fig:den_prof_rho03} and \ref{fig:den_prof_rho05}, the
mechanical failure occurs in the regions with largest spatial extent
of local density deviations from its average value. As can be
observed in Figs.\,\ref{fig:den_prof_rho03} and
\ref{fig:den_prof_rho05}, a dip in $\langle \rho \rangle_{s}(x)$
starts to develop within the regions with low average densities. The
process of local density decrease in these regions is accompanied by
simultaneous densification in the neighboring parts of the system,
similar to the constant volume loading
conditions~\cite{Priezjev18t}. Also, the shapes of density patterns
are largely preserved in the parts of the systems outside of the
low-density region of largest extent. Specifically, the density
patterns repeat themselves with increasing strain, the only
difference being their magnitudes and lateral shifts. The observed
increase in density magnitudes is due to rearrangement of material
when it flows from the low-density region to the regions of elevated
densities. This type of behavior has also been observed in porous
glasses loaded at constant volume~\cite{Priezjev18t}.

\vskip 0.05in


In contrast, at higher average glass densities, shown in
Fig.\,\ref{fig:den_prof_rho07}, the tensile loading at constant
pressure leads to qualitatively different evolution of the density
profiles.  In this case, the deformation results in gradual closure
of the existing pores. Specifically, the density profiles reveal a
gradual increase in density throughout the sample in the range of
intermediate strains. At large strains, the density profiles become
nearly flat outside of the region with largest extent of deviation
from the average density. Note, however, a small dip of about 10\%
remains at strain $\varepsilon_{xx}=1.6$. This behavior differs from
the one observed for loading at constant volume, where the dip in
the large-extent low-density region develops with increasing
strain~\cite{Priezjev18t}. These differences are due to a
competition between two effects. One of the effects is the material
rearrangement within the system and the other is the volume change
due to relaxation at constant pressure. The latter leads to a
decrease in total volume, which does not occur in systems undergoing
evolution at fixed volume.

\section{Conclusions}

In summary, dynamical evolution of porous glasses from a constant
volume state to zero-pressure state and mechanical response
properties of glasses at constant pressure were investigated using
atomistic simulations. We found that the transition to the
zero-pressure state is accompanied by significant increase in
average density and corresponds to a transition to lower energy
states. The temporal evolution of the pore distribution functions
during structural relaxation at zero pressure were examined. It was
revealed that a transition to configurations with a narrower
pore-size distributions is energetically favorable as compared to
the larger pores supported by built-in tensile stresses in the
systems. Both constant-volume and constant pressure configurations,
in the absence of deformation, show randomly distributed, isolated
ensembles of pores at higher average glass densities, while
interconnected porous structure is formed at lower densities. At
constant pressure, the equilibrium configurations consist of
narrower distributions of pores, which preserve, however, the major
characteristics of pore-size distribution functions observed for
samples equilibrated at constant volume.

\vskip 0.05in

The mechanical response properties, studied at zero pressure, reveal
a number of noteworthy features.  First, we found the exponent in
the power-law scaling of the elastic modulus with density is the
same as in the case of tensile, compressive, and shear testing of
the systems at constant volume. Consequently, the effect of built-in
pressures on elastic properties manifests itself primarily in the
corresponding porosity changes. Moreover, as our studies show, the
mechanical loading of porous systems at constant pressure differs
from that at fixed volume by the effect of total volume change. The
evolutions of pore ensembles in the systems subjected to tension, is
similar to the ones characteristic for the constant-volume loadings
at smaller average densities. In these cases, pore growth and
coalescence is favored as compared to the changes in the system
volume. At higher densities, however, the total volume changes
become the dominant effect in the response to the external loading.
This manifests itself in distinctly different behavior of density
profiles in the regime of higher densities as compared to the less
dense samples.

\section*{Acknowledgments}

Financial support from the National Science Foundation (CNS-1531923)
is gratefully acknowledged.  The article was prepared within the
framework of the Basic Research Program at the National Research
University Higher School of Economics (HSE) and supported within the
framework of a subsidy by the Russian Academic Excellence Project
`5-100'.  The molecular dynamics simulations were performed using
the LAMMPS numerical code~\cite{Lammps}. The distributions of pore
sizes were computed using the open-source software ZEO++ developed
at the Lawrence Berkeley National
Laboratory~\cite{Haranczyk12c,Haranczyk12,Haranczyk17}.
Computational work in support of this research was performed at
Wright State University's High Performance Computing Facility and
the Ohio Supercomputer Center.


%

%

%

%

%


%

%

%

%

%

%

%

\bibliographystyle{prsty}

\end{document}